\documentclass[12pt,twoside]{article}
\usepackage{amssymb}
\usepackage{amsmath}
\usepackage[makeroom]{cancel}
\usepackage{latexsym}
\usepackage{longtable}
\usepackage{epsfig}
\usepackage{dsfont}
\usepackage{graphicx,bbm,psfrag}
\graphicspath{{images/}}
\usepackage[ascii]{inputenc}
\numberwithin{equation}{section}

\begin{document}
\vspace*{2cm}
\begin{center}
 \LARGE{Interacting and Noninteracting Integrable Systems}\bigskip
 \end{center}
\begin{center}
\large{Herbert Spohn}
 \end{center}
 \begin{center}
Department of Mathematics, Columbia University, New York\\ 
Zentrum Mathematik and Physik Department, TUM,\\
Boltzmannstra{\ss}e 3, 85747 Garching, Germany.
 \tt{spohn@tum.de}
 \end{center}
\vspace*{4cm}
\textbf{Abstract}. We propose that the distinction between interacting and noninteracting integrable systems is 
characterized by the Onsager matrix. It being zero is the defining property of a noninteracting integrable system. 
To support our view  various classical and  quantum integrable chains are discussed.
\vspace*{6cm}
\begin{flushright} 6.12.2017
\end{flushright}

\newpage
\section{Introduction}\label{sec1}
Integrable, either classical or quantum, many-body systems require a highly fine-tuned hamiltonian. Except for trivial cases
the model has to be one-dimensional, typically with nearest neighbor interactions. Secondly every finite  $N$-site chain  
with periodic boundary conditions
is required to be integrable. For a classical system one employs the conventional definition: For a system with $N$ degrees of freedom
there exists a family of $N$ smooth functions on phase space, $\{I_1,...,I_N\}$, such that their range spans an $N$-dimensional domain 
and the Poisson brackets $\{I_m,I_n\} =0$. By convention $I_1$ is usually the total momentum and $I_2 = H$ is the hamiltonian generating the time evolution 
of the system. Hence the $I_m$'s are conserved.
Since the $I_m$'s are in involution, they can be used as new canonical coordinates. The corresponding canonically conjugate variables are the angle variables $
\{\theta_1,...,\theta_N\}$. The classical phase space then foliates into invariant $N$-tori, labelled by $\vec{I}$, with quasi-periodic motion of the angles as 
$\vec{\theta}(t) = \vec{\theta}(0) + \vec{\omega}(\vec{I})t\,\,\mathrm{mod}\,\, 2 \pi$. However, the naive quantum generalization of integrability turns out to be  not so helpful,
since all spectral projections of the hamiltonian are conserved.

In our context, we will use a different approach which is based on local conservation laws. Our notion works directly with the infinitely extended system and applies 
both to classical and quantum models. While general, to be sufficiently concrete let us consider the case of a spin-$\tfrac{1}{2}$ chain
with spin operators $\{\vec{\sigma}_j, j\in \mathbb{Z}\}$. A \textit{local} operator depends only on a finite number of spins, while a \textit{quasi-local} operator has exponentially decaying tails.
For example, the operator
\begin{equation}\label{1.1}
A_0 = \sum_{j \in \mathbb{Z}} c_j\vec{\sigma}_j\cdot\vec{\sigma}_{j+2}
\end{equation}
is local, if the coefficients $c_j=0$ except for a finite number of sites and quasi-local if $|c_j| \leq c \exp[-\gamma |j|]$ with some constants $c, \gamma > 0$.  The operator 
$A_0$ shifted by $x \in \mathbb{Z}$ reads
\begin{equation}\label{1.2}
A_x = \sum_{j \in \mathbb{Z}} c_j\vec{\sigma}_{j+x}\cdot\vec{\sigma}_{j+2+ x}.
\end{equation}
We now consider a spin chain for which the hamiltonian, $H$, has a local density and equally so for the charge $Q$,
\begin{equation}\label{1.3}
H = \sum_{x \in \mathbb{Z}} H_x, \quad Q = \sum_{x \in \mathbb{Z}} Q_x.
\end{equation}
Here $H_0$ is a local operator and $H_x$ denotes  $H_0$ when shifted by $x$.  
By definition, the charge $Q$ is \textit{locally conserved}, if there exist a current density operator $J_0$ such that 
\begin{equation}\label{1.4}
\mathrm{i}[H,Q_x] -J_x +J_{x+1} = 0.
\end{equation}
Summing over $x$, the telescoping sum vanishes and $Q$ itself  is conserved. 
Note that since $H$ and $Q$ have local densities, so do their commutator and the associated current. In the standard examples the density of $H$ has range 2
and the range is increased by one in the natural ordering of conserved charges.
In general, one would have to admit 
a $Q$-density with a unit cell larger than 1. We call a system \textit{integrable}, if there are infinitely many locally conserved charges. 
According to the available evidence there is no natural spin chain with, say, 47 locally conserved charges. Why there is such a strict dichotomy,
either a few or infinitely many conserved charges, stands as a deep and difficult puzzle.

From studying specific examples, it turns out that local is a too restrictive notion. One should allow  quasi-local densities and hence quasi-local current densities.

Going through the list of integrable chains, our definition agrees with the common usage, see the books and reviews 
\cite{T12,FT87,KBI93,G03,EFGKK05,rev1,SPA11,EF16,VR16}. Only the  requirement of conserved charges being
in involution has to be dropped. In fact, for quantum spin chains this was never considered to be natural condition. For example, in a model invariant under spin rotations,
the three components of the total spin are conserved and do not commute with each other. Amongst the locally conserved charges there is an Abelian subset with pairwise zero commutator. But in addition there can be  further non-commuting charges, then usually referred to as \textit{non-Abelian} \cite{F16}.

An exceptional case is the Calogero-Moser model for classical particles moving on the real line with hamiltonian
\begin{equation}\label{1.5}
H_\mathrm{CM} = \sum_{j=1}^N\tfrac{1}{2}p_j^2 + \tfrac{1}{2} \sum_{i \neq j = 1}^N\frac{1}{(q_i - q_j)^2}.
\end{equation}
The total momentum, $P$, is conserved with density
\begin{equation}\label{1.6}
P(x) = \sum_{j=1}^N\delta(q_j-x)p_j 
\end{equation}
and has the current density
\begin{equation}\label{1.7}
 \mathcal{J}_P(x) = \sum_{j=1}^N\delta(q_j-x)p_j^2 +  \tfrac{1}{2}\sum_{i \neq j = 1}^N(q_i - q_j)^{-3} \big(\chi(\{q_i < x < q_j\})- \chi(\{q_j < x < q_i\})\big), 
 \end{equation}
 where $\chi(\{\cdot\}) =1$ if the condition holds and $\chi(\{\cdot\}) =0$ otherwise. Because of the slow decay of the interaction potential the current is not quasi-local, 
 equally so for the energy density and densities of further conserved charges. Thus it remains to be seen whether the concepts developed here still apply to the long-ranged 
 Calogero-Moser model.
 
 In the more recent literature one notes the distinction between \textit{noninteracting} and \textit{interacting} integrable systems. There seems to be a general agreement to which class a given system belongs. For example the Lieb-Liniger $\delta$-Bose gas is interacting, so is the anisotropic XXZ Heisenberg chain,
but the ideal gas and the XY model are noninteracting. From a  classical perspective such a distinction is surprising. After all, according to the above discussion,
up to a coordinate transformation, all integrable systems are alike. 
In the quantum regime, a frequently used definition is to call  a model interacting if it is Bethe solvable and noninteracting if a mapping to free fermions or similar free theories
can be achieved. A related picture is based on quasi-particles. In noninteracting models they move independently, while interacting integrable models have non-trivial two-body scattering. While most likely both criteria properly capture the distinction, I suggest here a more physically motivated characterization. My proposal is fairly obvious. For a fluid in  three dimensions, if there is no interaction between particles, only ballistic transport is possible. To model dissipation requires adding short range interactions which then lead to non-zero viscosities and thermal conductivity.
Of course, a fluid has only five conservation laws. But still we may try to extend such a distinction between  ballistic and dissipative transport to one-dimensional integrable systems. For this purpose, it is assumed that initially  the system is in a spatially homogeneous generalized Gibbs ensemble (GGE) and one imposes an initial perturbation localized close to the origin. For  a
noninteracting integrable system the perturbation travels ballistically forever, however with dispersion since the velocity depends nonlinearly on the values of the conserved fields.
On the other hand for an interacting integrable system, on top there is dissipation leading to a strictly positive entropy production.

Before discussing specific models, let us explain a more precise formulation of our criterion. We will use a slightly symbolic notation so
to focus on  the main feature. But the spin model from above would be one example. Let us start from a one-dimensional lattice model, sites labeled by $j \in \mathbb{Z}$, 
a nearest neighbor hamiltonian, and the
conserved charges $Q^{(n)}$, $n = 0,1,...$\,. They have quasi-local densities as
\begin{equation}\label{1.8}
Q^{(n)} = \sum_{j \in \mathbb{Z}} Q^{(n)}_j.
\end{equation}
From the charges one constructs the generalized Gibbs ensemble (GGE)
\begin{equation}\label{1.9}
\frac{1}{Z}\exp\Big[ \sum_{n=1}^\infty \mu_n  Q^{(n)}\Big],
\end{equation}
where $\vec{\mu}$ is the vector of generalized chemical potentials.  Averages with respect to GGE are denoted by $\langle \cdot\rangle_{\vec{\mu}}$. 
If obvious from the context, the index $\vec{\mu}$ will be omitted. The object of interest is the correlator for the conserved charges,
\begin{equation}\label{1.10}
S_{mn}(j,t) = \langle Q^{(m)}_j(t)Q^{(n)}_0(0)\rangle_{\vec{\mu}}^\mathrm{c},
\end{equation}
where the time-evolved densities are denoted by $Q^{(n)}_j(t)$ and ${}^\mathrm{c}$ denotes connected truncation, i.e. the second cumulant. 
For such purpose we consider the conservation laws
\begin{equation}\label{1.11}
\frac{d}{dt}Q^{(n)}_j(t) -J^{(n)}_j(t) + J^{(n)}_{j+1}(t) = 0,
\end{equation}
where $J^{(n)}_j(t)$ denotes the current of $Q^{(n)}$ across the bond $(j-1,j)$. We also introduce the total current correlation function
\begin{equation}\label{1.12}
\Gamma_{mn}(t) = \sum_{j \in \mathbb{Z}} \langle J^{(m)}_j(t)J^{(n)}_0(0)\rangle_{\vec{\mu}}^\mathrm{c}.
\end{equation}
By a Lieb-Robinson bound and the good spatial mixing properties of the GGE, the summand has an exponential decay in $j$.
$S$ and $\Gamma$ are related by  the sum rule
\begin{equation}\label{1.13}
\sum_{j \in \mathbb{Z}} j^2 \big(S_{mn}(j,t) - S_{mn}(j,0)\big) = \int_0^t ds \int_0^tds'\, \Gamma_{mn}(s-s').
\end{equation}
Somewhat less known is a first order sum rule, which states that
\begin{equation}\label{1.14}
\sum_{j \in \mathbb{Z}} |j| \big(S_{mn}(j,t) - S_{mn}(j,0)\big)=  \int_0^t ds \int_0^tds'  \langle J^{(m)}(0,s)J^{(n)}(0,s')\rangle_{\vec{\mu}}^\mathrm{c},
\end{equation}
see \cite{MS15a} for a discussion.

In general, $\Gamma_{mn}(t)$ does not decay to $0$. The limit 
\begin{equation}\label{1.15}
\lim_{t \to \infty} \Gamma_{mn}(t) = D_{mn} ,
\end{equation}
defines the Drude weight matrix $D$. In principle, one should distinguish $\Gamma_{mn}(\infty)$ and $\Gamma_{mn}(-\infty)$. 
In the generality so far, the two limits could be different. By time-stationarity $\Gamma_{mn}(\infty) = \Gamma_{nm}(-\infty)$,
in particular $\Gamma_{mm}(\infty) = \Gamma_{mm}(-\infty) \geq 0$. In physical models, the charges, and hence their currents
are either even or odd under time-reversal. Then  $\Gamma_{mn}(\infty) = \Gamma_{mn}(-\infty)$ and $D_{mn} = D_{nm}$, see \cite{MS15a}. 
The Drude weight  $D$ is a positive-semidefinite symmetric matrix. We also define the Onsager matrix
\begin{equation}\label{1.16}
L_{mn} = \int_{\mathbb{R}} dt \big(\Gamma_{mn}(t)- D_{mn}\big).
\end{equation}
Since $L_{mn}$ comes from a covariance, $L$ is a symmetric matrix with non-negative eigenvalues. It could be that $L_{mn} = \infty$,
meaning that when integrating only over the interval $[-t,t]$ one arrives at a power law divergence as $t \to \infty$.   
 This is usually referred to as super-diffusive and requires a separate discussion, which is outside our present contribution.
 We propose to call a model \textit{noninteracting} if $L = 0$ as a matrix. Otherwise the model is \textit{interacting}. 
 
 The correlator has a time-independent normalization, since by the conservation law
   \begin{equation}\label{1.17}
\sum_{j \in \mathbb{Z}} S_{mn}(j,t) = \sum_{j \in \mathbb{Z}}  S_{mn}(j,0) = C_{mn},  
\end{equation}
 which is the static GGE susceptibility. The transport coefficients, called here generalized viscosities $\nu$, obtained by measuring  the spreading of the normalized correlator
 are defined through
  \begin{equation}\label{1.18}
\nu_{mn} = (LC^{-1})_{mn}.  
\end{equation}
 The physical interpretation becomes more transparent when considering a fixed charge $m=n$. Then for long times
 \begin{equation}\label{1.19}
\sum_{j \in \mathbb{Z}} j^2 S_{nn}(j,t) \simeq D_{nn} t^2 + L_{nn} t.
\end{equation}
$D_{nn} > 0$ signals that $S_{nn}(j,t)$ has a ballistic component. If   $L_{nn} = 0$, the correlator of the $n$-th conserved charge would spread ballistically,  linear in $t$, forever.
Otherwise there is a diffusive  $\sqrt {t}$ correction. In practice it might be difficult to observe such a sub-leading correction. 
As will be discussed, successful numerical computations of the viscosity cleverly focus on  a physical set-up for which $D_{nn} = 0$.

There are not so many models for which our proposition can be checked. A complete picture is available only for the hard rod fluid, 
briefly outlined in Section 5. For it $L$ has only one zero eigenvalue. The hard rod fluid is integrable and interacting. In Section 2
we establish that the XY chain in an external magnetic field satisfies indeed $L=0$. From our perspective the best understood interacting quantum model is the XXZ quantum Heisenberg chain, which will be discussed in Section 3, while Section 4 deals with the integrable classical  Faddeev-Takhtajan spin chain.
 
 
\section{The XY model in a transverse field}\label{sec2}
The XY model in a transverse field is a noninteracting integrable model and we want to understand in more detail whether and how the viscosity condition,
$L=0$, is satisfied. The model is chosen because the required input is readily available \cite{F16}. We mostly follow the notation there.  The hamiltonian of the spin chain
reads
\begin{equation}\label{2.1}
H_{\gamma,h} = \sum_{j \in \mathbb{Z}} \big( (1 + \gamma) \sigma_j^x \sigma_{j+1}^x + (1-\gamma)\sigma_j^y \sigma_{j+1}^y + 2h \sigma_j^z\big). 
\end{equation}
Compared to \cite{F16} we fix the coupling constant at $J=4$. $\gamma$ is the XY anisotropy and $h$ the strength of the transverse field.  $\vec{\sigma}_j$
are the Pauli matrices at lattice site $j$.  For the computations it is convenient to use 
Majorana fermions, denoted by $a_\ell$, which are symmetric operators satisfying the anti-commutation relations $\{a_\ell,a_{\ell'}\} = 2 \delta_{\ell\ell'}$, in particular 
$(a_\ell)^2 = 1$. The hamiltonian, the conserved charges, and their corresponding currents are all quadratic forms in the $a_\ell$'s, thus can be written as
\begin{equation}\label{2.2}
\mathrm{i} \sum_{\ell,\ell'} \mathcal{A}_{\ell\ell'} a_\ell a_{\ell'},
\end{equation}
where $\mathcal{A}$ is a real, anti-symmetric matrix, $\mathcal{A} = - \mathcal{A}^\mathrm{T}$. In addition, for these operators  the matrix $\mathcal{A}$ is invariant under a shift by 2. Hence after Fourier transform, lattice index to momentum $p \in [-\pi,\pi]$, $\mathrm{i} \mathcal{A}$ is represented by $p$-dependent $2\times 2$ matrix, the symbol of the operator. The symbol is written as linear combination of $ 
\mathds{1},\sigma_x,\sigma_y,\sigma_z$, using lower index to distinguish from the spins. For example, $H_{\gamma,h}$ is written in terms of Majorana fermions as
\begin{equation}\label{2.3}
\mathrm{i}  \sum_{\ell\in \mathbb{Z}} \big( - (1+\gamma) a_{2\ell}a_{2\ell +1}  + (1-\gamma)  a_{2\ell-1}a_{2\ell +2}  
 - 2ha_{2\ell+1} a_{2\ell+2}\big).
\end{equation}
The lower triangle of $\mathcal{A}$ follows from anti-symmetry, consistent with the anti-commutation relations.
Thus the corresponding symbol equals
\begin{equation}\label{2.4}
\hat{h}^{(\gamma,h)}(\mathrm{e}^{\mathrm{i}p}) =  \gamma \sin p \,\sigma_x + (h - \cos p)\sigma_y.
\end{equation}
 
The conservation laws of $H_{\gamma,h}$ come in two families: $I^{(n,+)}$ and $I^{(n,-)}$ with $n \geq 0$. Their symbols are
\begin{equation}\label{2.5}
\hat{i}^{(n,+)}(\mathrm{e}^{\mathrm{i}p})= \cos(np)\hat{h}^{(\gamma,h)}(\mathrm{e}^{\mathrm{i}p}), \quad \hat{i}^{(n,-)}(\mathrm{e}^{\mathrm{i}p})
= 4 \sin((n+1)p) \mathds{1}.
\end{equation}
The corresponding currents are denoted by $J^{(n,\pm)}$.  The currents $J^{(n,+)}$ are linear combination of conserved charges, thus time-independent. Therefore the  current correlations $\Gamma_{(m,\pm)(n,\pm)}(t) = D_{(m,\pm)(n,\pm)}$, except for the $(-,-)$ matrix element. On the other hand,
the $(n,-)$-currents do not commute with $H_{\gamma,h}$ and are given by
 \begin{equation}\label{2.6}
J^{(n,-)} =  - 2 \mathrm{i} \sum_{\ell \in \mathbb{Z}}\big( (1 +\gamma)(a_{2\ell}a_{2\ell  +2n+3} + a_{2\ell -1} a_{2\ell +2n} ) 
- (1-\gamma)  (a_{2\ell}a_{2\ell +2n -1} + 
a_{2\ell -1}a_{2\ell + 2n+4})  \big),
\end{equation}
which yields the symbol
\begin{equation}\label{2.7}
\hat{j}^{(n,-)}(\mathrm{e}^{\mathrm{i}p}) = 4 \sin((n+1)p)\big(\gamma \cos p\, \sigma_x + \sin p \,\sigma_y\big).
\end{equation}
The symbol of the time-dependent current, $\hat{j}^{(n,-)}(\mathrm{e}^{\mathrm{i}p},t)$, is obtained through
\begin{equation}\label{2.7a}
\hat{j}^{(n,-)}(\mathrm{e}^{\mathrm{i}p},t) = \exp[\mathrm{i}\hat{h}^{(\gamma,h)}(\mathrm{e}^{\mathrm{i}p})t] \hat{j}^{(n,-)}(\mathrm{e}^{\mathrm{i}p})
\exp[-\mathrm{i}\hat{h}^{(\gamma,h)}(\mathrm{e}^{\mathrm{i}p})t] .
\end{equation}
If $\gamma = 0$, then $\hat{j}^{(n,-)}(\mathrm{e}^{\mathrm{i}p},t)$ is constant in time, hence $L=0$. We thus assume $\gamma>0$. 
If $h=0$, then  in addition to $\hat{i}^{(n,+)}$, $\hat{i}^{(n,-)}$ the spin chain has also two-shift invariant conserved charges \cite{F16}, a case which will
 have to be studied separately. Henceforth we set $h>0$. 
 
Let us set 
\begin{equation}\label{2.8}
\omega^2 = (h - \cos p)^2 + (\gamma \sin p)^2.
\end{equation}
Using the formula of Rodrigues one finds
\begin{eqnarray}\label{2.9}
&&\hspace{-30pt}\hat{j}^{(n,-)}(\mathrm{e}^{\mathrm{i}p},t) \nonumber\\ 
&&\hspace{-10pt}= 
4 \sin((n+1)p)\Big(\cos(2\omega t) (\gamma \cos p\, \sigma_x + \sin p\, \sigma_y)+ \sin(2\omega t) \omega^{-1} 
\gamma (h - \cos p) \sigma_z \nonumber\\ 
&&\hspace{0pt}
+ \,(1-\cos(2\omega t))\omega^{-2} \sin p \big((\gamma^2-1) \cos p +h\big)\big(\gamma \sin p\, \sigma_x + (h- \cos p )\sigma_y\big) \Big). 
\end{eqnarray}

The next step is to compute the GGE average, which amounts to an average over the product of two quadratic operators. The GGE density matrix is
\begin{equation}\label{2.10}
\rho_{\mathrm{G}}  = Z^{-1} \mathrm{e}^{Q}
\end{equation}
with $Q$ some linear combination of conserved charges. From \eqref{2.5} one concludes that the symbol $Q$  is of the form
\begin{equation}\label{2.11}
\hat{q}(\mathrm{e}^{\mathrm{i}p}) = g_+(p) \hat{h}^{(\gamma,h)}(\mathrm{e}^{\mathrm{i}p}) + g_-(p) \mathds{1}.
\end{equation}
Here $g_+,g_-$ are smooth real functions on the circle $[-\pi,\pi]$ such that  $g_+(p) = g_+(-p)$ and $g_-(p) = -g_-(-p)$.
For all $i,j$ the GGE correlator reads
\begin{equation}\label{2.11a}
\mathrm{tr}[\rho_{\mathrm{G}}a_ia_j] = \delta_{ij} + \Gamma_{ij},  
\end{equation}
where the antisymmetric matrix $\Gamma$ is defined by the symbol
\begin{equation}\label{2.12}
\hat{\Gamma} (\mathrm{e}^{\mathrm{i}p}) = \tanh\big(\tfrac{1}{2}\hat{q}(\mathrm{e}^{\mathrm{i}p})\big).
\end{equation}
Then, according to (2.19) of \cite{F16}, the GGE average is given by
\begin{equation}\label{2.13}
\mathrm{tr}\big[\rho_{\mathrm{G}} A_0\big] 
= \frac{1}{4} \frac{1}{2\pi} \int _{-\pi}^{\pi} dp \,\mathrm{tr}[ \hat{\Gamma} (\mathrm{e}^{\mathrm{i}p}) \hat{a}
(\mathrm{e}^{\mathrm{i}p})],
\end{equation}
where $A$ is a 2 shift invariant quadratic operator which has density $A_j$ and symbol $\hat{a}
(\mathrm{e}^{\mathrm{i}p})$. In the current correlation for $H_{\gamma,h}$ there appears a second quadratic operator, $B$, with the same properties as $A$. Using the Pfaffian form of Wick's theorem, see (2.15) of \cite{F16}, one finds  
\begin{eqnarray}\label{2.14}
&&\hspace{-20pt}
\sum_{j \in \mathbb{Z}}\big( \mathrm{tr}\big [\rho_{\mathrm{G}} A_jB_0\big] - \mathrm{tr}\big [\rho_{\mathrm{G}} A_0\big]\mathrm{tr}\big [\rho_{\mathrm{G}} B_0\big]\big) \nonumber\\
&&\hspace{40pt}= \frac{1}{2} \frac{1}{2\pi} \int _{-\pi}^{\pi} dp\, \mathrm{tr}\big[ \big(\mathds{1} + \hat{\Gamma} (\mathrm{e}^{\mathrm{i}p})\big) \hat{a}
(\mathrm{e}^{\mathrm{i}p})\big(\mathds{1} +\hat{\Gamma} (\mathrm{e}^{\mathrm{i}p})\big) \hat{b}
(\mathrm{e}^{\mathrm{i}p})\big].
\end{eqnarray}
We conclude that the time dependence of the current correlation can be written as
\begin{equation}\label{2.15}
\Gamma_{mn}(t) = \frac{1}{2} \frac{1}{2\pi} \int _{-\pi}^{\pi} dp\, \mathrm{tr}\big[\big(\mathds{1} + \hat{\Gamma} (\mathrm{e}^{\mathrm{i}p})\big) \hat{j}^{(m,-)}
(\mathrm{e}^{\mathrm{i}p},t)\big(\mathds{1} +\hat{\Gamma} (\mathrm{e}^{\mathrm{i}p})\big)  \hat{j}^{(n,-)}(\mathrm{e}^{\mathrm{i}p})\big].
\end{equation}

To discuss the resulting Onsager matrix, we first consider the case $h \neq 1$. Then the sign of $\omega$ can be chosen as $\omega(p) >0$. $\Gamma_{mn}$ has the generic form 
\begin{equation}\label{2.16}
\Gamma_{mn}(t) - D_{mn} =  \int _{-\pi}^{\pi} dp \big( f_+(p)\cos(\omega(p)t) + f_-(p)\sin(\omega(p)t)\big),
\end{equation}
where $f_\pm$ are some smooth functions on the circle. The long time decay of $\Gamma_{mn}(t) - D_{mn}$ is determined by the critical points of $\omega$, 
which have to satisfy
\begin{equation}\label{2.17}
 \big( h - (1 - \gamma^2) \cos p\big) \sin p = 0,
\end{equation}
implying either $p = 0,\pi$ or $p = \arccos(h/(1 -\gamma^2))$. For $p=0$, the integrands $f_\pm$ vanish as $p^2$
resulting in a decay as $t^{-\frac{3}{2}}$, the same for $p=\pi$. But from the critical point $p = \arccos(h/(1 -\gamma^2))$, the generic decay is only $t^{-\frac{1}{2}}$
with oscillations. For the time-integral one obtains
\begin{equation}\label{2.18}
L_{mn} = \lim_{\epsilon \to 0_+}\int_\mathbb{R} dt \mathrm{e}^{-\epsilon |t|}\big(\Gamma_{mn}(t) - D_{mn}\big) = \int _{-\pi}^{\pi} dp \delta(\omega(p)) f_+(p),
\end{equation}
which implies $L_{mn} = 0$ because $\omega$ is supported away from $0$. 

If $h=1$, $\omega(p) \simeq |p|$ for small $p$ with a critical point only at $p = \pi$,
which dominates the long time behavior as $t^{-\frac{3}{2}}$. The time integral \eqref{2.18} still vanishes, since  $f_\pm(p) \simeq p^2$ 
near $p=0$ according to  \eqref{2.15}.
 
 We conclude that the XY model in a transverse field is integrable also in our sense, as established for $h \neq 0$. The total current correlation is time-dependent 
 with an oscillatory decay which generically is so slow that the time integral in \eqref{2.18} should  be regarded as an improper integral. But the viscosity vanishes over the entire parameter range.
 One should check also other integrable models, but the same features are to be expected.
 
 
\section{The Heisenberg XXZ chain}\label{sec3}

The hamiltonian of the XXZ chain reads
\begin{equation}\label{3.1}
H= \sum_{j \in \mathbb{Z}} \big(\sigma_j^x \sigma_{j+1}^x + \sigma_j^y \sigma_{j+1}^y + \Delta \sigma_j^z\sigma_{j+1}^z\big). 
\end{equation}
$\Delta >0$ is the anisotropy parameter. $0 < \Delta <1$ corresponds to easy-plane, $\Delta >1$ to easy-axis, while $\Delta = 1$ is the isotropic Heisenberg model.
The magnetization   
\begin{equation}\label{3.2}
M = \sum_{j \in \mathbb{Z}} \sigma^z_j
\end{equation}
is conserved with associated spin current
\begin{equation}\label{3.3}
J = \sum_{j \in \mathbb{Z}} \big(\sigma_j^x \sigma_{j+1}^y - \sigma_j^y \sigma_{j+1}^x\big).
\end{equation}
We will focus only on this particular local conservation law, denoting by $D$ its Drude weight and by $L$ its Onsager coefficient.
Instead of an arbitrary GGE, the usual thermal state $Z^{-1}\exp(-\beta H)$ is considered.
This  Drude weight has been  studied in considerable detail, see \cite{P11,P13,IP13,CP17,IN17}. It is convenient to introduce the shorthand
\begin{equation}\label{3.4}
\langle M;Q \rangle = \sum_{j \in \mathbb{Z}} \big( \langle M_j Q_0 \rangle - \langle M_0 \rangle \langle Q_0 \rangle\big)
\end{equation}
with  thermal average $\langle \cdot \rangle $.  Of course, $\langle \cdot \rangle $ could also refer to a GGE and $M_j$
could be the spatial translates of some other quasi-local operator $M_0$, correpondingly for $Q_j$.
In particular, according to \eqref{1.10},
\begin{equation}\label{3.5}
C_{mn} = \langle Q^{(m)}; Q^{(n)}\rangle.
\end{equation}
According to the method of hydrodynamic projections, the Drude weight is given by
\begin{equation}\label{3.6}
D = \sum_{m,n\geq 0} \langle J; Q^{(m)}\rangle (C^{-1})_{mn} \langle Q^{(n)} ;J\rangle.
\end{equation}
The $Q^{(n)}$'s are the conserved charges of the XXZ model. Here, the important point is to sum over all conserved charges. If some charges are missing,
one obtains at least a lower bound for $D$. The tricky point is hidden behind ``all''. The most common conserved charges have a strictly local density. On general grounds also  quasi-local charges, having an exponentially localized density, should be included in the sum. For a long time it was believed that the XXZ chain has only local conserved charges, for which it is known that
$\langle Q^{(n)};J\rangle =0$. Hence  \eqref{3.6} would yield $D=0$. But an exact steady state, enforced by boundary Lindbladians, exhibits ballistic transport \cite{P11}.  As a consequence a family of quasi-local charges was discovered \cite{P13}. Including all these charges 
in  \eqref{3.6}, one finds that still $D = 0$ for $\Delta \geq 1$. However for $\Delta <1$, $D>0$ and $D(\Delta)$ is fractal-like nowhere-continuous  function \cite{IP13,IN17}.

For the Onsager coefficient a recent result is available \cite{MKP17}, which strongly supports that $L >0$ for $\Delta > 1$. We explain some details of the
argument, since it well illustrates the difficulties. We start from a finite ring $ j = -\ell,...,\ell$ with periodic boundary conditions. $H^{[\ell]}$ is the corresponding finite volume hamiltonian, see \eqref{3.1}, $J^{[\ell]}$ the total finite volume spin current, see \eqref{3.3}, and 
\begin{equation}\label{3.7}
M^{[\ell]} = \sum_{|j| \leq \ell} \sigma^z_j = {\sum_{|m| \leq \bar{\ell}}}' m P_{m}^{[\ell]}
\end{equation}
is the magnetization with $P_{m}^{[\ell]}$ the projection onto all eigenstates of $M^{[\ell]}$ with eigenvalue $m$. The prime at the sum reminds that $m$ is summed in units of 2
and $\bar{\ell} = 2\ell +1$. 
The state at fixed $m$ is given by $\langle \cdot P_{m}^{[\ell]}\rangle_\ell/\langle P_{m}^{[\ell]}\rangle_\ell = \langle \cdot \rangle_{m,\ell}$, while $\langle \cdot \rangle_\ell $ denotes the
thermal state at volume $\bar{\ell}$. We choose some finite volume operator $Q^{[\ell]}$, such that $[H^{[\ell]},Q^{[\ell]}] =0$. By Schwarz inequality, using that also $[H^{[\ell]},P_{m}^{[\ell]}] =0$,
\begin{eqnarray}\label{3.8}
&&\hspace{-30pt}\big|\langle J^{[\ell]} Q^{[\ell]} \rangle_{m,\ell}^\mathrm{c}\big|^2 = \Big| \frac{1}{t} \int_0^t \!ds \langle J^{[\ell]}(s) (Q^{[\ell]}  -  \langle Q^{[\ell]} \rangle_{m,\ell})\rangle_{m,\ell}\Big|^2\nonumber\\
&&\hspace{80pt}  \leq 
\frac{1}{t^2} \int_{0}^t \!ds \int_0^t \!ds'   \langle J^{[\ell]}(s) J^{[\ell]}(s')\rangle_{m,\ell} \langle (Q^{[\ell]}){^2}\rangle_{m,\ell}^\mathrm{c}.
\end{eqnarray}
Hence, summing over  $\langle P_{m}^{[\ell]}\rangle_\ell$ and using stationarity,
\begin{equation}\label{3.9}
{\sum_{|m| \leq \bar{\ell}}}'(\langle (Q^{[\ell]})^2\rangle_{m,\ell}^\mathrm{c})^{-1} \langle P_{m}^{[\ell]}\rangle_\ell\big|\langle J^{[\ell]} Q^{[\ell]} \rangle_{m,\ell}^\mathrm{c}\big|^2 \leq 
\frac{\bar{\ell}}{t}\int_{-t}^t ds \big(1 - t^{-1}|s|\big)\sum_{|j| \leq \ell} \Gamma^\ell_j(s).
\end{equation}
Here
\begin{equation}\label{3.9b}
\Gamma^\ell_j(s) = \langle J_{(j,j+1)}(s) J_{(0,1)}(0) \rangle_\ell
\end{equation}
is the local current correlation, in which 
\begin{equation}\label{3.9a}
J_{(j,j+1)} = \sigma_{j}^x\sigma_{j+1}^y - \sigma_{j}^y\sigma_{j+1}^x
\end{equation}
denotes the $z$-spin current across the bond $(j,j+1)$. 

We first discuss the right side of \eqref{3.9}. By a Lieb-Robinson bound,  $J_{(j,j+1)}(s)$ is a quasi-local observable. Hence the infinite volume limit exists,
 \begin{equation}\label{3.9c}
\lim_{\ell \to \infty} \langle J_{(j,j+1)}(s) J_{(0,1)}(0) \rangle_\ell = \langle J_{(j,j+1)}(s) J_{(0,1)}(0) \rangle = \Gamma_j(s).
\end{equation}
 Also there is a velocity $v$ such that outside the cone $\{|j|  < vt+c_0\}$, with a suitably large but fixed constant $c_0$,  time correlations decay
 exponentially. In particular, since $\langle J_{(j,j+1)}(s)  \rangle_\ell =0 $, $\Gamma_j(s)$ is ensured to decay exponentially in $j$. Hence the total correlation
\begin{equation}\label{3.9d}
\Gamma(s)  = \sum_{j\in \mathbb{Z}} \Gamma_j(s).
\end{equation}
 is also well-defined. Unfortunately about $\Gamma(s)$  itself one knows only little, except for the bound
 \begin{equation}\label{3.9e}
|\Gamma(s)| \leq \langle J;J\rangle,
\end{equation}
which follows from Schwarz inequality.
So far these are general properties valid for any one-dimensional spin chain with a strictly local energy density. For the total current correlation, physically one expects an 
asymptotic power law decay as $\Gamma(t) - D \simeq t^{-\alpha}$, $\alpha >0$, possibly with logarithmic factors. If $\alpha \leq 1$, then $L=\infty$, no lower bound is required.
Thus, in our context  it is reasonable to assume that there is an integrable function $\phi(s)$ such that  
\begin{equation}\label{3.10}
\big|\sum_{|j| \leq \ell}\Gamma^\ell_j(s) \big|\leq \phi(s),
\end{equation}
independent of $\ell$, provided $|s| \leq t =\ell/v$. For  larger $s$ the perturbation originating at 0 would have traveled around the ring and \eqref{3.10} no longer holds. 
Also  at this point $\Delta > 1$ has to be imposed. In case of a non-zero Drude weight the upper bound $\phi$ would not decay to zero. 
 By dominated convergence one then concludes that 
\begin{equation}\label{3.11}
\lim_{\ell \to \infty} \int_{-\ell/v}^{\ell/v} ds \big(1 - (\ell/v)^{-1}|s|\big)\sum_{|j| \leq \ell}  \Gamma^\ell_j(s) = L.
\end{equation}

To establish a lower bound for $L$, one needs to study the left side of \eqref{3.9}.
Specifically we now choose  a locally conserved charge $Q$, which  is denoted  by $Q^{[\ell]}$ when restricted to the  volume $\bar{\ell}$.  One-dimensional chains, with a finite-range interaction and at non-zero temperature, have a finite correlation length and one can use the usual formulas from statistical mechanics. Firstly
we note that
\begin{equation}\label{3.14}
{\sum_{|m| \leq \bar{\ell}}}'\langle P_{m}^{[\ell]}\rangle_\ell = 1, \quad {\sum_{|m| \leq  \bar{\ell}}}'m\langle P_{m}^{[\ell]}\rangle_\ell = 0, \quad {\sum_{|m| \leq  \bar{\ell}}}' m^2\langle  P_{m}^{[\ell]}\rangle_\ell = \langle (M^{[\ell]})^2\rangle_\ell^\mathrm{c}.
\end{equation}
Hence
\begin{equation}\label{3.15}
\langle P_{m}^{[\ell]}\rangle_\ell \simeq \frac{2}{\sqrt{2\pi \kappa \bar{\ell}}}\,\mathrm{e}^{-{m^2/2\kappa}\bar{\ell}}
\end{equation}
with $\kappa = \langle M;M\rangle$, which implies that for $\langle (Q^{[\ell]})^2\rangle_{m,\ell}^\mathrm{c}$ and $\langle J^{[\ell]} Q^{[\ell]} \rangle_{m,\ell}^\mathrm{c}$ the range can be restricted to $|m| /\ell \ll 1$.
In particular, in that range 
\begin{equation}\label{3.12}
\lim_{\ell \to \infty} \frac{1}{\bar{\ell}} \langle (Q_m^{[\ell]})^2\rangle_{m,\ell}^\mathrm{c} =\langle Q;Q\rangle >0
\end{equation}
and one has to still study
\begin{equation}\label{3.13}
\frac{1}{\bar{\ell}}{\sum_{|m| \leq \bar{\ell}}}' \langle P_{m}^{[\ell]}\rangle_\ell\big|\langle J^{[\ell]} Q^{[\ell]} \rangle_{m,\ell}^\mathrm{c}\big|^2.
\end{equation}

In view of the equivalence of ensembles, we introduce the state
\begin{equation}\label{3.16}
\langle \cdot\rangle_{\ell, h} = \frac{1}{Z} \big\langle \cdot \prod_{|j| \leq \ell} \mathrm{e}^{-h \sigma_j^z}\big\rangle.
\end{equation}
Then
\begin{equation}\label{3.17}
\lim_{\ell \to \infty} \frac{1}{\bar{\ell}} \langle J^{[\ell]} Q^{[\ell]}\rangle_{\ell, h}^\mathrm{c} = \langle J;Q\rangle_h = g(h).
\end{equation}
$ g$ is a smooth function with $g(0) = 0$, since $\langle J^{[\ell]} \rangle_{\ell, h=0} =0$. For the first derivative one finds
\begin{equation}\label{3.18}
g'(0) = \langle J;Q;M\rangle,
\end{equation}
where 
\begin{equation}\label{3.18a}
\langle J;Q;M\rangle = \sum_{i,j \in \mathbb{Z}}  \langle J_{(i,i+1)} Q_j \sigma_0^z \rangle^\mathrm{c},
\end{equation}
the right hand average referring to the third cumulant. Hence
\begin{equation}\label{3.19}
\langle J^{[\ell]} Q^{[\ell]}\rangle_{m,\ell}^\mathrm{c} \simeq \bar{\ell}\frac{m}{\ell}  \langle J;Q;M\rangle \langle M;M\rangle^{-1}.
\end{equation}
Inserting \eqref{3.15} and  \eqref{3.19} in \eqref{3.13}, the factors of $\ell$ exactly balance and one concludes the lower bound 
\begin{equation}\label{3.20}
L \geq \frac{2}{v} \frac{\langle J;Q;M\rangle^2}{ \langle M;M\rangle  \langle Q;Q\rangle}.
\end{equation}
Repeating the argument for many charges, one obtains
\begin{equation}\label{3.20a}
L \geq \frac{2}{v}\frac{1}{ \langle M;M\rangle} \sum_{m,n\geq 0} \langle J;Q^{(m)};M\rangle (C^{-1})_{mn} \langle M;Q^{(n)}; J\rangle,
\end{equation}
compare with \eqref{3.6}.

One still has to make sure that $\langle J;Q;M\rangle$ does not vanish. A more complete discussion can be found in \cite{MKP17}. 
Here we simply choose the first conserved charge beyond the energy, which has the density
\begin{equation}\label{3.21}
Q_j = \sigma_{j-1}^x\sigma_{j}^z\sigma_{j+1}^y- \sigma_{j-1}^y\sigma_{j}^z\sigma_{j+1}^x -\Delta\big( \sigma_{j-1}^zJ_{(j,j+1)}
+ J_{(j-1,j)}\sigma_{j+1}^z\big),
\end{equation}
see \cite{BCDF16}. To have a proof of principle,
we only carry out the simplest case of $\beta = 0$, for which
\begin{equation}\label{3.22}
\langle M;M\rangle = 1, \hspace{3pt} \langle Q;Q\rangle = 2(1+ 2\Delta^2), \hspace{3pt}  g(h) = -4\Delta\langle \sigma_z\rangle_h\big(1 - 
\langle \sigma_z\rangle_h^2\big),\hspace{3pt}  \langle J;Q;M\rangle = 4 \Delta.
\end{equation}

An interesting control check has been undertaken in \cite{LZP17}. The initial state is domain wall with $\beta = 0$, a magnetic field $h>0$
on the right half-lattice, and a field $-h$ on the left half-lattice. Since $D=0$, on the Euler time scale the jump in the magnetization at the origin would stay put. But on the diffusive time 
scale the step broadens as an error function, while the current has a Gaussian profile. This allows one to determine the Onsager coefficient, which is
found to diverge as $\Delta \to 1_+$. At $\Delta =1$ measured is the time-integrated current across the origin, which is found to diverge as $t^{0.67}$. According to the sum rule \eqref{1.11}
the $|j|$-moment of $S$ diverges as $t^{0.67}$ and hence the $j^2$-moment  of $S$ as $t^{1.34}$. From the sum rule \eqref{1.6} one infers that $\Gamma(t) \simeq t^{-0.66}$, thus $L=\infty$ at $\Delta = 1$.
Since $D> 0$ for $\Delta < 1$, in this range  the method is no longer applicable.  
In view of the results reported in the next section, I conjecture that $L >0$ away from $\Delta = 1$ and $L \to \infty$ as $\Delta \to 1_\pm$. 
As pointed out in \cite{MMK17}, the dynamics becomes qualitatively different in the limit $h \to \infty$. 
 
\section{Classical integrable spin chains}\label{sec4}
The Toda chain is the most celebrated integrable classical chain which is build from particles indexed by the  one-dimensional lattice and coupled through nonlinear springs
with hamiltonian
\begin{equation}\label{4.1}
H_{\mathrm{Toda}} = \sum_{j \in \mathbb{Z}}\big( \tfrac{1}{2} p_j^2 + V(q_{j +1} - q_{j})\big), \quad V(x) = \mathrm{e}^{-ax}.
\end{equation}
While time-correlations of the conserved fields have been studied through molecular dynamics \cite{KD16}, the particular issue of dissipative corrections remains unexplored.
Less studied are integrable classical spin models. In this case one considers a spin lattice with $\vec{S}_j  \in \mathbb{R}^3$ under the constraint
$|\vec{S}_j| = 1$. The interaction is nearest neighbor with energy
\begin{equation}\label{4.2}
H =  \sum_{j \in \mathbb{Z}} h(\vec{S}_j,\vec{S}_{j+1}).
\end{equation}
The time evolution is governed by 
\begin{equation}\label{4.3}
\frac{d}{dt} \vec{S}_{j} = \vec{\nabla}_{\! j} H \wedge \vec{S}_j,
\end{equation}
where $\vec{\nabla}_{\! j}$ denotes differentiation with respect to $\vec{S}_j$. The dynamics is hamiltonian
with canonical coordinates 
\begin{equation}\label{4.3b}
\phi_j = \arctan(S_{j,2}/S_{j,1}), \quad s_j = S_{j,3},
\end{equation}
and suitable boundary conditions at $\phi_j = \pm\pi$, resp. $s_j = \pm 1$. For the Landau-Lifshitz chain, the interaction is quadratic
\begin{equation}\label{4.3a}
 h_\mathrm{LL}(\vec{S}_j,\vec{S}_{j+1}) = S_{j,1}S_{j+1,1} + S_{j,2}S_{j+1,2} + \Delta S_{j,3}S_{j+1,3},
\end{equation}
which can be viewed as the large spin limit of the XXZ chain. The Landau-Lifshitz chain is not integrable. 
As observed by Faddeev and Takhtajan \cite{FT87}, the model 
becomes integrable for the special choice
\begin{eqnarray}\label{4.4}
&&\hspace{-40pt}h(\vec{S},\vec{S}')  = \log \big| \cosh(\rho S_3) \cosh(\rho S_3') \nonumber\\
&&\hspace{20pt}+ \coth^2(\rho) \sinh(\rho S_3) \sinh(\rho S_3') + \sinh^{-2}(\rho) F(S_3) F(S_3')
(S_1S_1' + S_2S_2')\big|
\end{eqnarray}
with 
\begin{equation}\label{4.5}
F(S) = \big((\sinh^{2}(\rho) - \sinh^{2}(\rho S))/(1 - S^2)\big)^{\frac{1}{2}}.
\end{equation}
$\rho$ is the only parameter of the model, either real or purely imaginary. For our discussion it is more convenient to choose the real parameter $\delta = \rho^2$. 
$\delta$ measures the anisotropy. $\delta = 0$ is the isotropy point, while $\delta >0$ corresponds to easy-axis (mostly motion along the 3-axis)
and $\delta <0$ to easy-plane (mostly motion in the 1-2 plane). 

The $3$-component of the spin is locally conserved. Its current correlation in thermal equilibrium has been studied  through molecular dynamics \cite{PZ13} 
for a system size up to $5000$ sites and at an inverse temperature $\beta = 1/4$. For $\delta =1$ the Drude weight vanishes  and the Onsager coefficient
 $L = 0.38$
with an apparently quickly decaying  $\Gamma(t)$. In contrast to the XXZ model, for $\delta = -1$ the current correlation can simulated without any additional  effort. The Drude weight is non-zero ($D = 0.61$ in the simulation) and $\Gamma(t) - D$ is positive with rapid decay. This strongly indicates
that the Onsager coefficient is strictly positive. At $\delta = 0$, the point of isotropy, one finds through a direct simulation that  $\Gamma(t) \propto t^{-0.65}$.
These findings suggest an Onsager coefficient $L(\delta) >0$, but diverging as $\delta \to \pm 0$. Such qualitative phase diagram seems to be identical to the one of the XXZ model, in fact with the same anomaly exponent at the transition point.

 
\section{Hard rod fluid}\label{sec5}
We consider a one-dimensional classical fluid of hard rods. The hard rods have length $a>0$, positions $q_j\in\mathbb{R}$, and velocities $v_j\in \mathbb{R}$. They move freely except for elastic collisions conserving momentum and energy, whereby two hard rods  exchange their velocities upon contact.  In a collision, we label the particles so as to maintain their velocities. Thus particle $j$ moves along a straight line, $\dot{q}_j = v_j$, interrupted by jumps back and forth of size $a$ due to collisions. Clearly, such a system with $N$ rods has $N$ conservation laws labeled by their velocities. 

The hard rod fluid is  the only interacting integrable system which is  in a certain sense completely  understood, including the dissipative corrections
which result from the random-like jumps by $\pm a$. Thus the hard rods serve as a
sufficiently simple model to which more complicated models can be compared. We indicate only a few items of interest in the context of our discussion. More details are available in \cite{DS17,DS17a}.   

We study directly the infinitely extended system. In a GGE the velocities are independent with the common probability  density function $h(v)dv$, which is assumed to be smooth. But our formulas would extend also to a discrete set of delta functions. On the other hand positions are correlated due to the hard core repulsion.
The particle density is denoted by $\rho$, $0 < a\rho < 1$. 
For the infinite system it is known that in the class of sufficiently regular measures, the only time-stationary measures are given by a GGE 
\cite{dobrods}.

For hard rods the correlator of the conserved fields has been computed exactly  for a general GGE \cite{LPS68}. Thus one also knows the total current correlation function 
\begin{eqnarray}\label{5.1}
&&\hspace{10pt}\Gamma_{vv'}(t) = \delta(t) (a\rho)^2 (1-a\rho)^{-1}\big(\delta(v-v') r(v)h(v)  - |v-v'|h(v)h(v')\big)\nonumber\\[1 ex]
&&\hspace{20pt}+\rho(1-a\rho)^{-2}\big(\delta(v - v')v^2h(v) - a\rho (v^2 + v'^2) h(v)h(v')  + (a\rho)^2 d_2h(v)h(v') \big),\nonumber\\
&&\hspace{20pt}
\end{eqnarray}
with the abbreviations
\begin{equation}\label{5.2}
r(v) = \int_\mathbb{R} dw h(w)|w-v|, \quad d_2 = \int_\mathbb{R}  dw h(w)w^2 
\end{equation}
and assuming that $\int_\mathbb{R} dwh(w)w = 0$ \cite{S82,S91}. The second term on the right in \eqref{5.1} is the Drude weight $D_{vv'}$, which is symmetric. Integrating against the test function $\psi$ one obtains
\begin{equation}\label{5.2a}
\int_{\mathbb{R}^2}dvdv'\psi(v)D_{vv'}\psi(v') = \rho (1-a\rho)^{-2} \int_{\mathbb{R}}dvh(v)v^2\Big(\psi(v) - a\rho \int_{\mathbb{R}}dv'h(v')\psi(v')\Big)^2
\end{equation}
implying $D >0$ as an operator. Since $D$ is a finite rank perturbation of a multiplication operator, its spectrum is purely continuous and consists of $\mathbb{R}_+$.

The first term on the right of  \eqref{5.1} is proportional to $\delta(t)$.
Its integral is then the Onsager matrix given by 
\begin{equation}\label{5.3}
L_{vv'} = (a\rho)^2 (1-a\rho)^{-1}\big(\delta(v-v') r(v)h(v)  - |v-v'|h(v)h(v')\big).
\end{equation}
Since the susceptibility is given by
\begin{equation}\label{5.4}
C_{vv'} = \rho\big(\delta(v-v')h(v) + a\rho (a\rho - 2) h(v)h(v')\big),
\end{equation}
one obtains for the viscosity 
\begin{equation}\label{5.5}
\nu_{vv'} = a(a\rho) (1-a\rho)^{-1}\big(\delta(v-v') r(v) - h(v)|v-v'|\big).
\end{equation}
Clearly the hard rod fluid is interacting and becomes noninteracting in the ideal gas limit $a \to 0$. 

The Onsager matrix has a single zero eigenvalue with $f(v) = 1$ as eigenfunction. Physically this corresponds to the density, whose current is the itself locally conserved momentum, hence no dissipation for this special mode. All other eigenvalues of $L$ are separated by a spectral gap from $0$. Since $D>0$, in contrast to the XXY and Fadeev-Takhtajan model, ballistic transport cannot be turned off by making a  particular  
choice of the conserved field and model parameters. On the other hand, on the basis of the hydrodynamic equations including the nonlinear Navier-Stokes correction,  one can determine the entropy production \cite{DS17}, which in the quadratic approximation is proportional to the viscosity $\nu_{vv'}$. Thus, although the model is integrable, there is still the connection between dissipation and entropy production, as well known from the theory of fluids.
 
\section{Conclusions and outlook}\label{sec6}
On the classical side it would be of interest to investigate in more detail the Toda lattice, in particular to find out about its Onsager matrix through molecular dynamics. On the quantum side, spectacular progress has been achieved in identifying the Euler type hydrodynamics for several interacting integrable chains
\cite{BCDF16,CDY16,bvkm2,DYZ17,A1,PDCBF17,ID17}. The Onsager matrix is difficult to access, however, both theoretically and through DMRG simulations. Only if the Drude weight vanishes, one seems to have sharp tools, compare with Section 3. A nonvanishing Drude weight poses the serious problem to subtract a dominating background.

While this is not the place to enter into details, if one moves to the realm of non-integrable chains, the behavior changes drastically. In many models the Drude weight vanishes and for  the correlator of the conserved charges one observes diffusive spreading of a non-moving central peak. Exceptions may result from almost conserved charges \cite{MS15,S16}.
For classical anharmonic chains momentum is conserved provided the interaction depends only on the relative distance of particles, i.e.
no pinning potential. Then in addition to the central peak there will be two sharp sound peaks with non-zero speed which generically spread super-diffusively,
see \cite{S16} for a review.  The same behavior is found for a one-dimensional classical fluid with short range interactions. Since one relies on hydrodynamic arguments, there is every reason to predict that also one-dimensional quantum fluids show such characteristics. On the other hand, for quantum lattice models momentum conservation is broken. Currently it is
an open problem whether there is some other conserved charge which would play the role of the momentum for a classical anharmonic chain.
To rephrase, one searches for a non-integrable chain  whose Drude weight does not vanish.\\\\
\textbf{Acknowledgements}. I am grateful for most useful comments by  B. Doyon, M. Fagotti, E. Ilievski, V. Mastropietro, J. De Nardis, and T. Prosen on a first draft of my notes.

\end{document}